\RequirePackage{fix-cm}
\documentclass[twocolumn,pdftex]{svjour3}          
\smartqed  
\usepackage{graphicx}
\usepackage{amssymb,amsfonts,amsmath}
\usepackage{dsfont}
\usepackage{natbib}
\usepackage{tikz}

\DeclareMathOperator*{\Ind}{\mathds{1}}

\newcommand{\ABCmu}{ABC$\mu$}
\newcommand{\smcABC}{sisABC}
\newcommand{\smcABCmu}{sis\ABCmu}

\newcommand{\mhABCmu}{mh\ABCmu}

\newcommand{\Sset}{\mathbb{S}}
\newcommand{\R}{\mathbb{R}}
\newcommand{\N}{\mathbb{N}}
\newcommand{\E}{\mathbb{E}}
\newcommand{\m}{M}
\newcommand{\abcK}{\kappa}
\newcommand{\beps}[1][]{{\varepsilon^{#1}_{1:K}}}
\newcommand{\lkl}{f(x_0|\theta)}
\newcommand{\simlkl}{f(x|\theta)}
\newcommand{\simulkl}{x\sim f(\cdot\:|\theta)}
\newcommand{\abctarget}{\pi_{\tau}(\theta,x|x_0)}
\newcommand{\abcmargtarget}{\pi_{\tau}(\theta|x_0)}
\newcommand{\abcmumargtarget}{\pi_{\tau}(\beps|x_0)}
\newcommand{\abcmutarget}{\pi_{\tau}(\theta,\beps|x_0)}

\newcommand{\AVGND}{$\overline{\text{ND}}$}
\newcommand{\AVGNDm}{\overline{\text{ND}}}
\newcommand{\WR}{WR}
\newcommand{\DIA}{DIA}
\newcommand{\AVGCC}{CC}
\newcommand{\BOX}{BOX}
\newcommand{\FRAG}{FRAG}
\newcommand{\CONN}{CONN}
\newcommand{\ODBOX}{OD\BOX}
\newcommand{\MWIDTH}{M-EXPL}
\newcommand{\CFDPK}{CDF-$\Delta$PK}
\newcommand{\ACFDPK}{ACF-$\Delta$PK}
\newcommand{\CPK}{CDF-PK}
\newcommand{\MATTR}{M-ATTR}

\providecommand{\abs}[1]{\lvert#1\rvert}	

\newenvironment{singlespaceddescription}{
\begin{description}
  \setlength{\itemsep}{1pt}
  \setlength{\parskip}{0pt}
  \setlength{\parsep}{0pt}
}{\end{description}}

\newcommand{\pDiv}{\delta_{\text{Div}}}
\newcommand{\pAttach}{\delta_{\text{A}}}
\newcommand{\rDup}{\lambda_{\text{Dup}}}
\newcommand{\rLnkAdd}{\lambda_{\text{Add}}}
\newcommand{\rLnkDel}{\lambda_{\text{Del}}}

\RequirePackage[normalem]{ulem} 
\RequirePackage{color}\definecolor{RED}{rgb}{1,0,0}\definecolor{BLUE}{rgb}{0,0,1} 
 
\begin{document}

\title{Monte Carlo algorithms for\\ model assessment via conflicting summaries
\thanks{OR is partly financially supported by the Wellcome Trust (fellowship WT092311), the National Science Foundation (grant NSF-EF-08-27416); PP and CR by the French Agence Nationale de la Recherche (grant ANR-09-BLAN-0145-01 'EMILE'); and SR by a Royal Society Wolfson Merit award and the MRC-HPA Centre on Environment and Health.}
}


\author{Oliver Ratmann \and Pierre Pudlo \and Sylvia Richardson \and Christian Robert}


\institute{O. Ratmann \at
              Biology Department, Duke University, Box 90338 Durham, NC 27708, USA\\
              \email{oliver.ratmann@duke.edu}           
	   \and
	   P. Pudlo \at
           Institut de Math\'ematiques et de Mod\'elisation de Montpellier, Universit\'e Montpellier 2,
 	   Montpellier, France
           \and
           S. Richardson \at
           Department of Epidemiology and Biostatistics, Imperial College London, London, United Kingdom
           \and
           C. Robert \at
           Universit\'e Paris-Dauphine, Paris, France\\ 
           \email{xian@ceremade.dauphine.fr}           
}

\date{Received: date / Accepted: date}

\maketitle

\begin{abstract} The development of statistical methods and numerical algorithms for model choice is vital to
many real-world applications. In practice, the ABC approach can be instrumental for sequential model design;
however, the theoretical basis of its use has been questioned. We present a measure-theoretic framework for
using the ABC error towards model choice and describe how easily existing rejection, Metropolis-Hastings and
sequential importance sampling ABC algorithms are extended for the purpose of model checking. We considering a
panel of applications from evolutionary biology to dynamic systems, and discuss the choice of summaries, which 
differs from standard ABC approaches. The methods and algorithms presented here may provide the
workhorse machinery for an exploratory approach to ABC model choice, particularly as the application of
standard Bayesian tools can prove impossible.  
\keywords{Approximate Bayesian Computation \and Metropolis-Hastings \and Sequential Monte Carlo \and model choice}
\end{abstract}

\section{Introduction}\label{intro}
Approximate Bayesian Computation (ABC) methods have appeared in the past ten years as a way to handle
intractable likelihoods and posterior densities \begin{equation}\label{e:posterior} \pi(\theta|x_0)\propto \lkl
\pi(\theta) \end{equation} that arise under high dimensional, data-generating models. For example, complex
coalescent models are known to generate, currently, latent structures that are too high dimensional to bring a
reliable numerical approximation in practical computer time. Originally developped for population genetics, ABC
has since been applied to many applied problems where Bayesian analysis has long been contemplated
but previously remained elusive \citep[]{Beaumont2010, Pudlo2011}. 

As with other approximation methods like variational Bayes \citep[]{Jaakkola2000,MacKay2002} or indirect
inference \citep[]{Heggland2004}, ABC suffers from a limited ability to quantify the uncertainty in the
approximation of the posterior \eqref{e:posterior}. Moreover, the loss of information brought by the ABC
approximation implies that the application of parts of standard Bayesian machinery, such as the Bayes factor,
is fraught with difficulties \citep[]{Robert2011}.

While much of Bayesian model checking is based on evaluating model predictions, ABC uses such model predictions
for parameter inference. In this perspective, it is natural to attempt using the pseudo-data $x$ that is
generated by ABC Monte Carlo algorithms both for parameter inference and model assessment. There is no issue of
bias in doing so because we are considering a simulation technique rather than an inferential method: the data
itself is only ``used once". Some of us have called for some technical refinement of this framework, called ABC
under model uncertainty (\ABCmu) \citep[]{Robert2009PNAS}, particularly as this technique can generate further
insight in practice \citep[]{Drovandi2011, Ratmann2010}. In opposition to more formal Bayesian model choice
approaches \citep{Toni2008,grelaud2009}, a key to the
validation of ABC as a model assessment in 
\cite{Ratmann2009} relies on the very fact that likelihood
computations and comparisons under a model can be replaced by assessing the amount of fit between
simulations $x$ from that model and the observed data $x_0$.

In this paper, we first present technical modifications that reflect our consensus view on ABC under model
uncertainty. We then describe basic, yet efficient Metropolis-Hastings and sequential importance sampling ABC
algorithms for approximate parameter inference and model checking, which forms the main contribution of this
paper.  On purpose, these algorithms are closely related to existing, popular ABC algorithms to show how easily
these
methods can be extended to incorporate model checking at no or little additional computational cost.  These
algorithms are presented in Section \ref{Xtd:ABC}, following a description of their theoretical foundations in
Section \ref{s:section1}. We illustrate these algorithms on a panel of applications, ranging from population
genetics to network evolution and dynamic systems (Sections 4-6). We discuss the relative advantages of both
algorithms, and compare these to a hybrid algorithm that seeks to combine the strengths of either method. Given
the difficulties associated with using approximate Bayes factors for model choice, we conclude that the
algorithms presented in this paper may provide the workhorse machinery for a viable, exploratory approach to
model choice when the likelihood is computationally intractable (Section \ref{s:discuss}).

\section{A measure-theoretic framework for ABC}\label{s:section1}
To formalize the setting of ABC-led inference with an application to diagnostic model assessment in mind, we first present a
measure-theoretic framework that updates the previous formulation in \cite{Ratmann2009}. The extended ABC algorithms in Section \ref{Xtd:ABC} also handle goodness-of-fit type analyses and follow immediately from this re-interpretation of ABC.

\subsection{The ABC approach} \label{s:abc_intro}
As in most ABC settings, we suppose that pseudo-data $x\in\mathcal{X}$ can be efficiently simulated for any
vector of model parameters $\theta\in\Theta$ from a data-generating process $f$ that defines, perhaps
implicitly, the likelihood. We also consider a set of $K$ summaries $\Sset=\big\{S_1,\dotsc,S_K\big\}$, a real-valued
distance function $\rho$, and the one-dimensional ABC kernel 
\begin{equation*}
\abcK(e;\tau)=1/\tau\Ind\big\{\abs{e}\leq \tau/2\big\}
\end{equation*}
with tolerance $\tau>0$. To circumvent likelihood evaluations, \cite{Pritchard1999} first proposed the
rejection sampler rejABC in Table~\ref{t:rejectionsamplers}A.

The target density of rejABC on the augmented space $\Theta\times\mathcal{X}$ is therefore
\begin{equation}\label{e:AbcTargetDensity}
\abctarget\propto     \abcK\Big(\rho\big(\Sset(x),\Sset(x_0)\big);\tau\Big)       \simlkl           \pi(\theta).
\end{equation}
In typical applications, the auxiliary variable $x$ is extremely high-dimensional, and lower-dimensional
summaries are used to compare the simulated data with the observed data. The realized error $\varepsilon=
\rho\big(\Sset(x),\Sset(x_0)\big)$ is then accepted by the ABC algorithm when within a prescribed tolerance $\tau$.
ABC is a valid, non-parametric estimation method in that, as $\tau\rightarrow 0$,
the marginal density $\abcmargtarget$ of \eqref{e:AbcTargetDensity} approaches the true posterior
distribution \eqref{e:posterior} if the summaries are sufficient for $\theta$ under the model. Otherwise,
\eqref{e:AbcTargetDensity} converges to the posterior distribution $\pi(\theta|\Sset(x_0))$ when
$\tau\rightarrow 0$. \cite{Fernhead2011} and \cite{Dean2011} show that an ABC-based inference is converging
(in the number of observations) if the parameter $\theta$ is identifiable in the distribution of $\Sset(x)$. 

\begin{table}[tbp]
	\begin{minipage}[t]{0.45\textwidth}
		\vspace{-0.2cm}
		\begin{minipage}[b]{0.001\textwidth}
			{\bf A}\newline\vspace{2.4cm}

		\end{minipage}
		\fbox{
		\begin{minipage}[b]{0.99\textwidth}	
{\bf Algorithm rejABC}\\ on $\Theta\times\mathcal{X}$ to sample from Eq.\thinspace\ref{e:AbcTargetDensity}:\\[-1mm]
\begin{singlespaceddescription}\label{s:stdabcmu}
\item[rejABC1] Sample $\theta\sim\pi(\theta)$, simulate $\simulkl$ and compute $\varepsilon=\rho\big(\Sset(x),\Sset(x_0)\big)$.
\item[rejABC2]
Accept $(\theta,x)$ with probability proportional to $\abcK(\varepsilon;\tau)$, and go to rejABC1.
\end{singlespaceddescription}
		\end{minipage}}
	\end{minipage}
	\hspace{0.5cm}	
	\begin{minipage}[t]{0.45\textwidth}
		\vspace{0.4cm}
		\begin{minipage}[b]{0.001\textwidth}
			{\bf B}\newline\vspace{2.4cm}

		\end{minipage}
		\fbox{
		\begin{minipage}[b]{0.99\textwidth}		
{\bf Algorithm rej\ABCmu}\\ on $\Theta\times\R^K$ to sample from Eq.\thinspace\ref{e:ABCmutargetdensity}:\\[-1mm]
\begin{singlespaceddescription}\label{s:stdabcmub}
\item[rej\ABCmu 1] Sample $\theta\sim\pi(\theta)$, simulate $x\sim f(\cdot\:|\theta)$ and compute $\varepsilon_k=\rho_k\big(S_k(x),S_k(x_0)\big)$, $k=1,\dotsc,K$.
\item[rej\ABCmu 2]
Accept $(\theta,\beps)$ with probability proportional to $\prod_k\abcK(\varepsilon_k;\tau_k)$, and go to rej\ABCmu 1.
\end{singlespaceddescription}
		\end{minipage}}
	\end{minipage}
	\vspace{0.4cm}
\caption{Rejection samplers for ABC and \ABCmu.}\label{t:rejectionsamplers}
\end{table}

\subsection{ABC on error space}
\label{s:abc_error}
The error $\varepsilon$ computed in algorithm rejABC (Table \ref{t:rejectionsamplers}A) is, in fact, a compound error term that may reflect both stochastic fluctuations in simulating from $f$ as well as
systematic biases between $f$ and the data. To exploit this information, we reformulate ABC~as providing simulations on the joint space of model parameters ($\theta$) and summary
errors ($\beps$). Algorithm rej\ABCmu~in Table \ref{t:rejectionsamplers}B uses the projection 
\begin{equation}\label{e:Kdimerror}
\xi_{x_0,\theta}\colon\mathcal{X}\to\R^K,\quad x\to\beps=(\varepsilon_1,\dotsc,\varepsilon_K),\\
\end{equation}
$\varepsilon_k=\rho_k\big(S_k(x),S_k(x_0)\big)$, which induces the image measure (abusively denoted by)
\begin{equation}\label{e:Kdimerrormeasure}
\begin{split}
&\xi_{x_0,\theta}(E_1\times\dotsc\times E_K)\\
&= \mathbb{P}_f\Big(\:\xi_{x_0}^{-1}(E_1\times\dotsc\times E_K)\:\Big|\:\theta\:\Big)\\
&=\int_{\xi_{x_0}^{-1}(E_1\times\dotsc\times E_K)}\:\:f(dx|\theta)
\end{split}
\end{equation}
on the associated Borel image $\sigma$-algebra, conditional on $x_0,\theta$.
The density of \eqref{e:Kdimerrormeasure} with respect to a suitable measure on the $K$-dimensional error space
will be denoted (again abusively) by
\begin{equation}\label{e:xiK}
\xi_{x_0,\theta}\colon\R^K\to\R^+_0,\quad\beps\to\xi_{x_0,\theta}(\beps)\,.
\end{equation}
This multi-dimensional error density is
thus the image of the sampling density $f(\cdot|\theta)$ by the transform $\xi_{x_0,\theta}$. It can be
interpreted as the prior predictive error density conditional on $\theta$. 

\begin{example} In many applications, pseudo-data $x$ is simulated on a finite space $\mathcal{X}$. Then,
$f(dx|\theta)$ is a counting measure, say  
\begin{equation*}
f(dx|\theta)= \sum_{i=1}^{N_x} f_i\delta_{x_i}(dx). 
\end{equation*}
Hence, the image measure $\xi_{x_0,\theta}(d\varepsilon)$ is again a counting measure, say 
\begin{equation*}
\xi_{x_0,\theta}(d\varepsilon)= \sum_{j=1}^{N_\varepsilon} \xi_j\delta_{\varepsilon_j}(d\varepsilon),
\end{equation*}
where $N_\varepsilon$ is the size of $\xi_{x_0,\theta}(\mathcal{X})$, $0<N_\varepsilon\leq N_x^2<\infty$. 
\end{example}

From a computational perspective, \eqref{e:xiK} is in practice intractable. In direct analogy to ABC, we
circumvent numerical evaluations of \eqref{e:xiK} through simulating from this density as illustrated in 
algorithm rej\ABCmu\ in Table~\ref{t:rejectionsamplers}B. The target density of rej\ABCmu\ on the augmented
space $\Theta\times\R^K$ is
\begin{equation}\label{e:ABCmutargetdensity}
\abcmutarget\propto     \prod_k \abcK\big(\varepsilon_k;\tau_k\big)       \xi_{x_0,\theta}(\beps)           \pi(\theta).
\end{equation}
By construction, the marginal target densities $\abcmargtarget$ of rejABC and rej\ABCmu\ coincide if $\abcK\big(\rho(\Sset(x),\Sset(x_0);\tau\big)$ in rejABC can be written, up to a constant of proportionality, as $\prod_k \abcK\big(\rho_k(S_k(x),S_k(x_0);\tau_k\big)$ (to be used in rej\ABCmu) for some choice of $\tau_k$; see the Appendix. For example, if ABC is run with the standard indicator kernel, this is equivalent to 
using the Manhattan distance
\begin{equation*}
\rho(\Sset(x),\Sset(x_0)=\max\big\{\rho_k\big(S_k(x),S_k(x_0)\big)\big\}
\end{equation*}
and $\tau_k=\tau$.
The utility of this reformulation was first discussed in \cite{Ratmann2009}: it is possible to relate the marginal density 
$\abcmumargtarget$ of \eqref{e:ABCmutargetdensity} to standard Bayesian error measures. The marginal ABC error
density can be understood as the prior predictive error density \citep[]{Box1980} that is 
re-weighted by error magnitude
\begin{equation}\label{e:ABCmuerrordensity}
\abcmumargtarget\propto \prod_k \abcK\big(\varepsilon_k;\tau_k\big)\pi_{x_0}(\beps).
\end{equation}

\section{Extending existing ABC algorithms}\label{Xtd:ABC}
Existing ABC algorithms are easily extended to sample from $\abcmutarget$ for the purpose of parameter inference and model assessment.

In Table \ref{t:mhsamplers}, we contrast the Metropolis-Hastings ABC sampler \citep[]{Marjoram2003} to its extension that samples from \eqref{e:ABCmutargetdensity}. To demonstrate the validity of algorithm \mhABCmu, let $z=(\theta,\beps)$ and note that, on the augmented space, the proposal density of mh\ABCmu\ is $q(z\to z^\prime)= \xi_{x_0,\theta^\prime}(\beps[\prime])q(\theta\to\theta^\prime)$. Therefore, detailed balance is satisfied precisely for $\abcmutarget$: 
\begin{equation*}
\begin{split}
&\frac{mh(z,z^\prime)}{mh(z^\prime,z)}\:=\quad\frac{q(\theta^\prime\rightarrow\theta)\:\pi(\theta^\prime)\prod_k\abcK(\varepsilon^\prime_k;\tau_k)}{q(\theta\rightarrow\theta^\prime)\:\pi(\theta)\prod_k\abcK(\varepsilon_k;\tau_k)}\\[1mm]
&\quad=\frac{q(z^\prime\rightarrow z)\:\pi(\theta^\prime)\:\prod_k\abcK(\varepsilon^\prime_k;\tau_k)\:\xi_{x_0,\theta^\prime}(\beps[\prime])}{q(z\rightarrow z^\prime)\:\pi(\theta)\:\prod_k\abcK(\varepsilon_k;\tau_k)\:\xi_{x_0,\theta}(\beps)}\\[1mm]
&\quad=\frac{q(z^\prime\rightarrow z)\:\pi_{\tau}(z^\prime|x_0)}{q(z\rightarrow z^\prime)\:\pi_{\tau}(z|x_0)}.
\end{split}
\end{equation*}
\cite{Bortot2007} proposed a Metropolis-Hastings sampler on the space $\Theta\times\mathcal{X}\times[0,\infty)$ for the purpose of parameter inference when the tolerance $\tau$ is by design a random variable. For clarity, we note that this algorithm requires an extra proposal density $q(\tau\to\tau^\prime)$, and has a target density different to both \eqref{e:AbcTargetDensity} and \eqref{e:ABCmutargetdensity}.

\begin{table}[tbp]
	\begin{minipage}[t]{0.45\textwidth}
		\begin{minipage}[b]{0.001\textwidth}
			{\bf A}\newline\vspace{5.8cm}

		\end{minipage}
		\fbox{
		\begin{minipage}[b]{0.99\textwidth}	
{\bf Algorithm mhABC}\\ on $\Theta\times\mathcal{X}$ to sample from Eq.\thinspace\ref{e:AbcTargetDensity}:\\[-1mm]

Set initial values $\theta^0$ and compute $x^0\sim f(\,\cdot\,|\theta^0)$.\\
\begin{singlespaceddescription}\label{s:mcmcabc}
\item[mhABC1] If now at $\theta$ propose a move to $\theta^\prime$ according to a proposal density $q(\theta\rightarrow\theta^\prime)$.
\item[mhABC2] Simulate $x^\prime\sim f(\cdot|\theta^\prime,\m)$ and compute $\varepsilon^\prime=\rho\big(\Sset(x^\prime),\Sset(x_0)\big)$.
\item[mhABC3] Accept $(\theta^\prime,x^\prime)$ with probability
\begin{equation*}
\begin{split}
&mh(\theta,x;\theta^\prime,x^\prime)=\\
&\quad\min\Bigg\{1\:,\frac{q(\theta^\prime\rightarrow\theta)}{q(\theta\rightarrow\theta^\prime)}\times\frac{\pi(\theta^\prime)\:\abcK(\varepsilon^\prime;\tau)}{\pi(\theta)\:\abcK(\varepsilon;\tau)}\Bigg\},
\end{split}
\end{equation*}
and otherwise stay at $(\theta,x)$. Return to mhABC1.
\end{singlespaceddescription}
		\end{minipage}}
	\end{minipage}
	\hspace{0.5cm}
	\begin{minipage}[t]{0.45\textwidth}
		\vspace{0.2cm}
		\begin{minipage}[b]{0.001\textwidth}
			{\bf B}\newline\vspace{6.6cm}

		\end{minipage}
		\fbox{
		\begin{minipage}[b]{0.99\textwidth}		
{\bf Algorithm mh\ABCmu}\\ on $\Theta\times\R^K$ to sample from Eq.\thinspace\ref{e:ABCmutargetdensity}:\\[4mm]
Set initial values $\theta^0$ and compute $\varepsilon_k^0=\rho_k\big(S_k(x^0),S_k(x_0)\big)$ where 
$x^0\sim f(\,\cdot\,|\theta^0)$.\\
\begin{singlespaceddescription}\label{s:mhabcmu}
\item[mh\ABCmu 1] If now at $\theta$ propose a move to $\theta^\prime$ according to a proposal density $q(\theta\rightarrow\theta^\prime)$.
\item[mh\ABCmu 2] Simulate $x^\prime\sim f(\cdot|\theta^\prime)$ and compute $\varepsilon^\prime_k=\rho_k\big(S_k(x^\prime),S_k(x_0)\big)$, $k=1,\dotsc,K$.
\item[mhABC 3] Accept $(\theta^\prime,\beps[\prime])$ with probability
\begin{equation*}
\begin{split}
&mh(\theta,\beps;\theta^\prime,\beps[\prime])=\\
&\quad\min\Bigg\{1\:,\frac{q(\theta^\prime\rightarrow\theta)}{q(\theta\rightarrow\theta^\prime)}\times\frac{\pi(\theta^\prime)\:\prod_k\abcK(\varepsilon_k^\prime;\tau_k)}{\pi(\theta)\:\prod_k\abcK(\varepsilon_k;\tau_k)}\Bigg\},
\end{split}
\end{equation*}
and otherwise stay at $(\theta,\beps)$. Return to mh\ABCmu 1.
\end{singlespaceddescription}		
		\end{minipage}}
	\end{minipage}
\caption{Vanilla Metropolis-Hastings samplers for ABC and \ABCmu. We include the summary error $\varepsilon$ in algorithm mhABC to emphasize that all the computations required for \ABCmu\ are already performed in the corresponding ABC algorithm.}\label{t:mhsamplers}
\end{table}

\begin{table*}[tbp]
	\begin{minipage}[t]{0.45\textwidth}
		\begin{minipage}[b]{0.001\textwidth}
			{\bf A}\newline\vspace{10.8cm}

		\end{minipage}
		\fbox{
		\begin{minipage}[b]{0.99\textwidth}	
{\bf Algorithm \smcABC}\\ on $\N\times\Theta\times\mathcal{X}$ to sample, marginally, from Eq.\thinspace\ref{e:AbcTargetDensity}:\\[2mm]
Set the initial particle system at $n=1$:  for $i=1,\dotsc,N$ compute $\theta^i_1\sim\pi(\theta)$, $x^i_1\sim f(\,\cdot\,|\theta^i_1)$, $\varepsilon^i_1=\rho\big(\Sset(x^i_1),\Sset(x_0)\big)$, and then, for $i=1,\dotsc,N$, $W^i_1=\abcK(\varepsilon^i_1;\tau_1)\:\big/\:\sum_{j=1}^N\abcK(\varepsilon^j_1;\tau_1)$.\\

For $n= 2,\dotsc,n^*$, do:\\
$\hphantom{X}$Set $i=1$, $c_n=  \abcK(0;\tau_n)$ and repeat:\\[-2mm]
\begin{singlespaceddescription}\label{s:smcabc}
\item[$\hphantom{X}${\bf\smcABC 1}] Propose the $i$th ancestor index $I^\prime$ from $i=1,\dotsc,N$ with probabilities $W^i_{n-1}$, $\theta^\prime\sim M_n(\theta^{I^\prime}_{n-1};\,\cdot\,)$, $x^\prime\sim f(\,\cdot\,|\theta^\prime)$ and compute $\varepsilon^\prime=\rho\big(\Sset(x^\prime),\Sset(x_0)\big)$.
\item[$\hphantom{X}${\bf\smcABC 2}] With probability $\abcK(\varepsilon^\prime;\tau_n)/c_n$,
set $(I^i_n,\theta^i_n,x^i_n)\leftarrow (I^\prime,\theta^\prime,x^\prime)$, compute the unnormalized weight
\begin{equation*}
w^i_n= \pi(\theta^i_n)\:\Big/\sum_{j=1}^N W^j_{n-1}M_n(\theta^j_{n-1};\theta^i_n)
\end{equation*}
and increment $i\leftarrow i+1$. If $i=N$, go to \smcABC 3. Else return to \smcABC 1.
\item[$\hphantom{X}${\bf\smcABC 3}] Compute the normalized weights
\begin{equation*}
W^i_n = w^i_n \:\Big/ \sum_{i=1}^Nw^i_n
\end{equation*}
and update $n\leftarrow n+1$. If $n<n^*$, go to \smcABC 1.
\end{singlespaceddescription}
		\end{minipage}}
	\end{minipage}
	\hspace{0.5cm}
	\begin{minipage}[t]{0.45\textwidth}
		\vspace{-11.35cm}
		\begin{minipage}[b]{0.001\textwidth}
			{\bf B}\newline\vspace{10.8cm}

		\end{minipage}
		\fbox{
		\begin{minipage}[b]{0.99\textwidth}		
{\bf Algorithm \smcABCmu}\\ on $\N\times\Theta\times\R^K$ to sample, marginally, from Eq.\thinspace\ref{e:ABCmutargetdensity}:\\[2mm]
Set the initial particle system at $n=1$:  for $i=1,\dotsc,N$ compute $\theta^i_1\sim\pi(\theta)$, $x^i_1\sim f(\,\cdot\,|\theta^i_1)$, $\varepsilon^i_{1k}=\rho_k\big(S_k(x^i_1),S_k(x_0)\big)$, and then, for $i=1,\dotsc,N$, $W^i_1=\prod_k\abcK(\varepsilon^i_{1k};\tau_{1k})\:\big/\:\sum_{j=1}^N\prod_k\abcK(\varepsilon^j_{1k};\tau_{1k})$.\\

For $n= 2,\dotsc,n^*$, do:\\
$\hphantom{X}$Set $i=1$, $c_n= \prod_k\abcK(0;\tau_{nk})$ and repeat:\\[-2mm]
\begin{singlespaceddescription}\label{s:smcabcmu}
\item[$\hphantom{X}${\bf\smcABCmu 1}] Propose the $i$th ancestor index $I^\prime$ from $i=1,\dotsc,N$ with probabilities $W^i_{n-1}$, $\theta^\prime\sim M_n(\theta^{I^\prime}_{n-1};\,\cdot\,)$, $x^\prime\sim f(\,\cdot\,|\theta^\prime)$ and compute $\varepsilon^\prime_k=\rho_k\big(S_k(x^\prime),S_k(x_0)\big)$ for all $k$.
\item[$\hphantom{X}${\bf\smcABCmu 2}] With probability $\prod_k\abcK(\varepsilon^\prime_k;\tau_{nk})/c_n$,
set $(I^i_n,\theta^i_n,\varepsilon^i_{n,1:K})\leftarrow (I^\prime,\theta^\prime,\beps[\prime])$, compute the unnormalized weight 
\begin{equation*}
w^i_n= \pi(\theta^i_n)\:\Big/\sum_{j=1}^N W^j_{n-1}M_n(\theta^j_{n-1};\theta^i_n)
\end{equation*}
and increment $i\leftarrow i+1$. If $i=N$, go to \smcABCmu 3. Else return to \smcABCmu 1.
\item[$\hphantom{X}${\bf\smcABCmu 3}] Compute the normalized weights
\begin{equation*}
W^i_n = w^i_n \:\Big/ \sum_{i=1}^Nw^i_n
\end{equation*}
and update $n\leftarrow n+1$. If $n<n^*$, go to \smcABCmu 1.
\end{singlespaceddescription}
		\end{minipage}}
	\end{minipage}
\caption{Vanilla sequential importance samplers for ABC and \ABCmu. Both algorithms require to specify a decreasing sequence of tolerances $\tau_n$  for $n=1,\dotsc,n^\star$. The first tolerance $\tau_1$ is here set large enough such that $W^i_1>0$ for all $i=1,\dotsc,N$, and subsequent ones can be set automatically \citep[]{Moral2008} or according to an annealing scheme. Using the indicator kernel, the acceptance probability in  \smcABC 2 (and \smcABCmu 2) is either zero or one \citep[]{Toni2008}. Typically, the variance of the proposal kernel $M_n$ is modified at each stage to improve the convergence of the algorithm \citep[]{Beaumont2009}.}\label{t:smcsamplers}
\end{table*}

In Table \ref{t:smcsamplers}, we contrast the popular sequential importance sampler (SIS) for ABC
\citep[]{Toni2008} to its extension that samples from \eqref{e:ABCmutargetdensity}.  To keep the particle
system at the $n$th stage alive, these algorithms augment the state space with an additional random variable
$I^i_n\in\{1,\dotsc,N\}$ .  This is the ancestor index of the particle at stage $n-1$ from which the $i$th
particle at stage $n$ is derived. Although we are only interested in the marginal target density of \smcABCmu\
on the space $\Theta\times\R^K$, the validation of \smcABCmu\ as a proper self-normalised importance sampler
with respect to $\abcmutarget$ proceeds as in \cite{Beaumont2009}. We note that at stage $n>1$, proposed
particles follow the law
$q_n(z_n,I_n)=\xi_{x_0,\theta_n}(\varepsilon_{n,1:K})M_n(\theta_n,\theta^{I_n}_{n-1})W^{I_n}_{n-1}$. However,
after integrating out the index $I_n$ and accounting for the weight correction, we have marginally for any
$\pi_\tau$-integrable function $h$ that
\begin{equation*}
\begin{split}
&\E^{q_n}\big(W_n h(z_n)\big)\\
&\propto\iint h(z_n)\sum_{j=1}^N W_n\:{\textstyle\prod_k}\abcK(\varepsilon_{n,k};\tau_{n,k})\:\times\\
&\hspace{4cm}q_n(z_n,j)\nu(z_{n-1})dz_{n-1}dz_n\\
&\propto\iint h(z_n)\:\pi(\theta_n)\:{\textstyle\prod_k}\abcK(\varepsilon_{n,k};\tau_{n,k})\:\xi_{x_0,\theta_n}(\varepsilon_{n,1:K})\:\times\\
&\hspace{4cm}\nu(z_{n-1})dz_{n-1}dz_n\\
&\propto\E^{\pi_\tau}\big(h(z_n)\big),
\end{split}
\end{equation*}
independently of the distribution of the previous $z_{n-1}$. The algorithm is therefore a proper importance
sampling scheme and the proposal kernel $M_n$ can be adapted to the previous $z_{n-1}$ \citep[]{Beaumont2009}.
We note that the proposed method of resampling ancestor indices is more generally known as sequential
importance sampling with Rao-Blackwellized rejection control, and has been successfully applied to a variety of
complex inference problems  \citep[]{Liu1998}.

\begin{figure*}[tbp]
	\centering
	\includegraphics[type=pdf,ext=.pdf,read=.pdf,width=\textwidth]{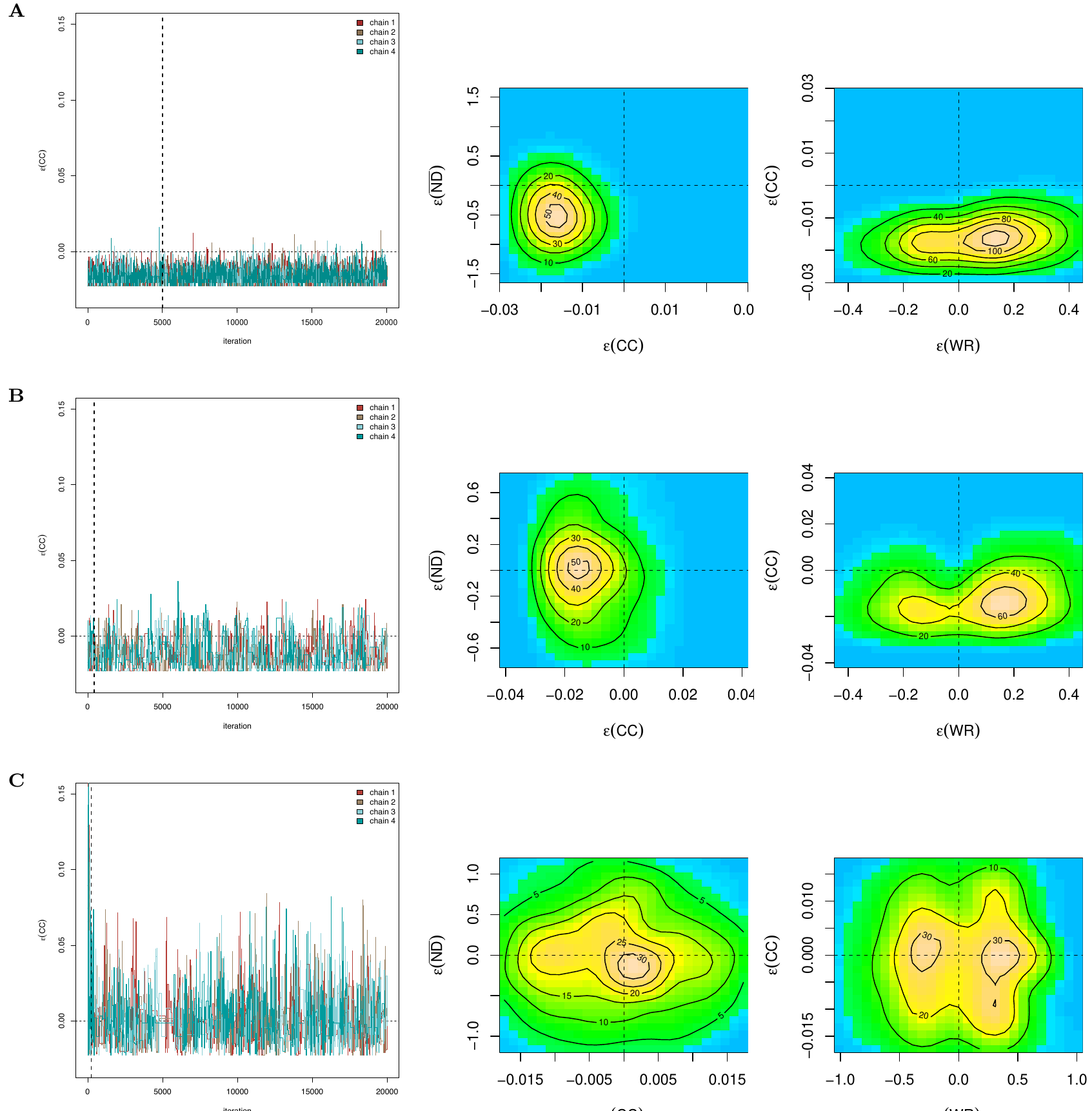}			
	\caption{Discrepancies between several network evolution models and protein network data from {\it T. pallidum}. We show results for two alternative evolution models and two different observation models, (A) DD+LNK+PA-L, (B) DD+PA-L, (C) DD+PA-BP; results for DD+LNK+PA-BP are similar to (A) and not shown. Left column: trajectories of the \AVGCC\ error for four Markov chains that are generated in parallel. Convergence was most difficult to achieve for the model in (A) and involved re-parameterization because the support of some parameters spanned several orders of magnitude. Right columns: two dimensional estimates of the seven dimensional ABC error density, as reproduced with average shifted histograms.}\label{f:networkevolutionmodels}      
\end{figure*}

\section{Application of \mhABCmu\ to network evolution}\label{s:netvolution}
We previously proposed a different ABC Metropolis-Hastings sampler to simultaneously fit and assess the adequacy of a model against the data \citep[]{Ratmann2009}. This algorithm combined the marginal densities $\xi_{x_0,\theta}(\varepsilon_k)= \int \xi_{x_0,\theta}(\beps) d\varepsilon_{-k}$ in an ad-hoc manner, thereby losing the true dependency structure between the ABC errors in \eqref{e:ABCmutargetdensity}. Marginally on error space, algorithms \mhABCmu\ and  \smcABCmu\ sample from the joint distribution of ABC errors. In this section, we re-visit the examples on which our original ideas were developped, and illustrate how iterative model design may benefit from the ability to sample from the joint error distribution $\abcmumargtarget$.

When large-scale, protein-protein interaction data became available, it came as a surprise that their topological features deviate markedly from those expected under standard random graphs. A variety of simple mathematical models were subsequently proposed to explain some of these unexpected topological features \citep[]{Stumpf2006a}. These models grow networks iteratively node by node until a given finite network size is reached. The preferential attachment model (PA) was able to reproduce, roughly, the observed fat-tailed empirical distribution of node degrees (number of outgoing edges of a node). Subsequently, more biologically plausible models based on the duplication of nodes and (local) link divergence among the duplicates (DD), as well as (global) link rewiring were proposed (LNK). We analyzed mixtures of these mechanisms on protein network data that was obtained with high-throughput bait-prey technologies \citep[]{Ratmann2009}. 
Briefly, the DD+PA mixture model has three parameters, and DD+LNK+PA has five parameters. Available network data reflects only a subset of the true interaction network. To interface the evolution models with data, several observation models have been formulated. Given a simulated complete network, we can randomly sample links until the observed number of links is matched (L). Alternatively, we can mimick the experimental design by randomly sampling bait and prey proteins until the observed numbers are matched, and record associated links (BP); see the Appendix for details. These sampling schemes do not add any parameters. All prior densities are uninformative. 

To compare observed network data $x_0$ to the corresponding simulations $x$, we use seven summary statistics that reflect local and global properties of the network topology: average node degree (\AVGND); within-reach distribution (\WR), the mean probability of how many nodes are reached from one node within distance $k=1,2,\dotsc$, where distance is the minimum number of edges that have to be visited to reach a node $j$ from node $i$;  diameter (\DIA), the longest minimum path among pairs of nodes in a connected component; cluster coefficient (\AVGCC), the mean probability that two neighbours of a node are themselves neighbours; fragmentation (\FRAG), the percentage of nodes not in the largest connected component; log connectivity distribution (\CONN), $\log\big(p(k_1,k_2)\AVGNDm^2\big)/\big(k_1p(k_1)k_2p(k_2)\big)$, the depletion or enrichment of edges ending in nodes of degree $k_1$, $k_2$ relative to the uncorrelated network with same node degree distribution $p(k)$; box degree distribution (\ODBOX), the probability distribution of boxes with $k$ edges to nodes outside the box. The choice of these summaries is discussed in \citep[]{Ratmann2007}.

We applied algorithm \mhABCmu\ in Table \ref{t:mhsamplers}B with annealing on the ABC thresholds and the variance of the Gaussian proposal density $q(\theta\rightarrow\theta^\prime)$ across a set of network evolution models. The sampler converged rapidly to the target density \eqref{e:ABCmutargetdensity}, as assessed by four parallel Markov chains that were started at overdispersed values. Markov chains were highly autocorrelated. Figure \ref{f:networkevolutionmodels} displays representative Markov chain trajectories, and two dimensional estimates of the seven dimensional ABC error density for three different models.

Inspection of the multi-dimensional ABC error density enabled us to diagnose specific deficiencies in models of network evolution. In our approach to ABC, each error $\varepsilon_k$ corresponds directly to one summary, thereby enabling targeted model refinement. For example, Figure \ref{f:networkevolutionmodels}A illustrates that the fitted link rewiring model with link subsampling (DD+LNK+PA-L) fails to match local connectivity patterns (\AVGCC) in the observed {\it Treponema pallidum} network \citep[]{Titz2008} while it reproduces global topological patterns (\AVGND, \WR). This prompted us to remove the link rewiring component and to consider DD+PA-L, which results in some improvement (Figure \ref{f:networkevolutionmodels}B). We also replaced link subsampling with bait-prey sampling. The fitted DD+PA-BP model is consistent with the seven considered aspects of the data in that the point $\beps=0$ is well in the support of the ABC error density (Figure \ref{f:networkevolutionmodels}C).

\begin{figure*}
	\centering
	\includegraphics[type=pdf,ext=.pdf,read=.pdf,width=\textwidth]{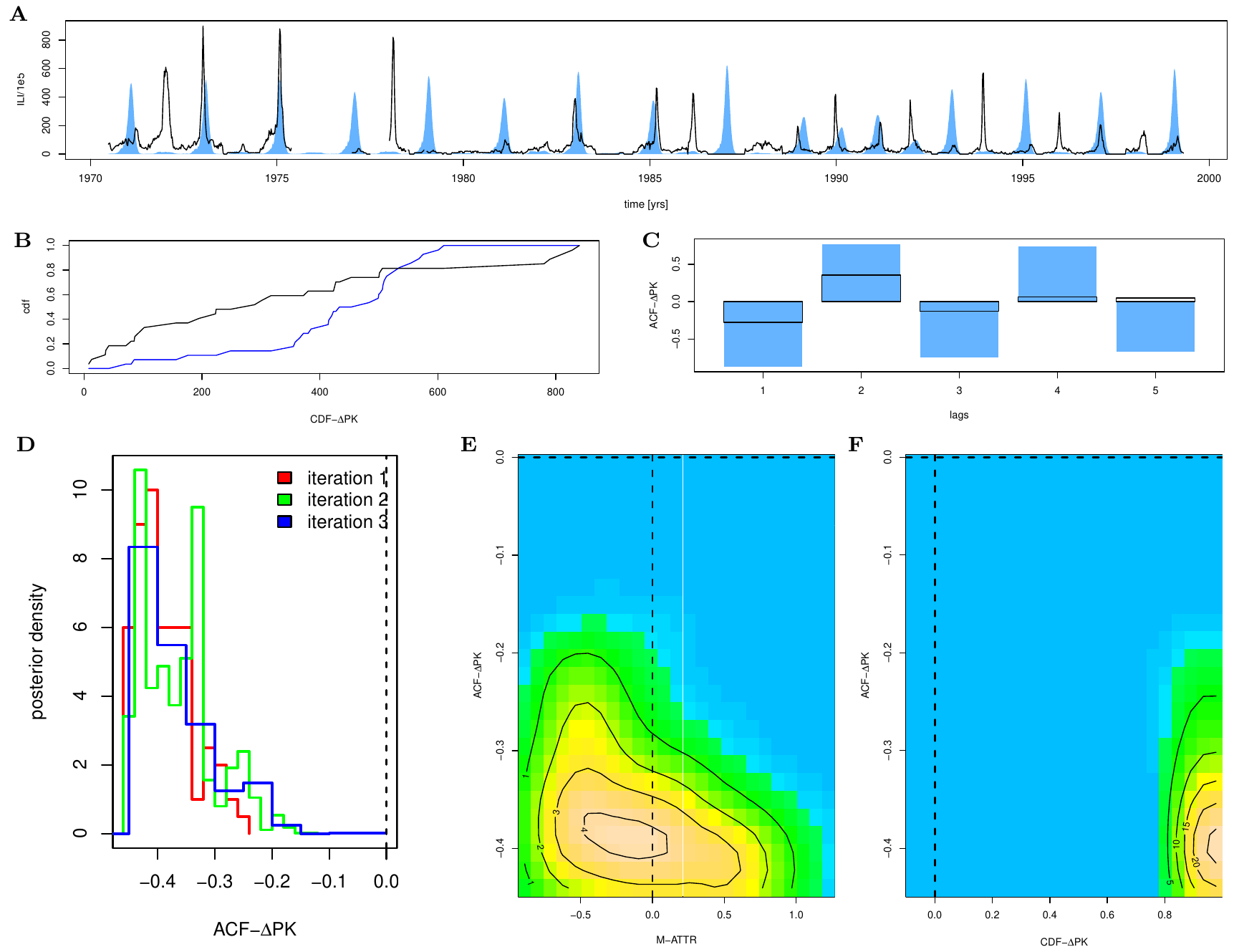}
	\caption{Discrepancies between the SIRS model (\ref{e:SIRS}-\ref{e:SIRSobs}) and aggregated influenza-like illness counts associated to influenza virus A (H3N2) from the Netherlands, 1970-99, as detected by \smcABCmu. Empirical data is shown in black, and simulations in blue; simulations correspond to one representative, accepted particle of  \smcABCmu. (A) Incidence data, standardized per 100,000 individuals. (B) Cumulative distribution function of differences between successive peaks. (C) Autocorrelations between successive peak differences. Simulated incidence data shows too strong and too regular inter-seasonal variation. Algorithm \smcABCmu\ was seeded with 100 samples from \mhABCmu\ that were obtained by thinning 500 iterations of 4 chains after burn-in, and then run for two more iterations with 1000 particles at the final ABC thresholds. (D) Histogram of the evolution of the \ACFDPK\ error across importance sampling iterations. (E-F) Two-dimensional estimates of the five-dimensional ABC error density. While the fitted model reproduces the magnitude of disease incidence (\MATTR), it does not match patterns of observed interannual seasonal variation (\CFDPK, \ACFDPK).}\label{f:SIRSdiscrepancies}      
\end{figure*}

\section{Application of \smcABCmu\ to dynamic systems}\label{s:dynamic}
The behavior of complex dynamic systems is often described by sets of non-linear ordinary or stochastic differential equations. Typically, these systems are only partially observed in aggregated form, and associated models are not analytically solvable and may fail to reproduce important aspects of the sytem's dynamics. \cite{Toni2008} proposed sequential ABC algorithms (in Table \ref{t:smcsamplers}A) for parameter inference in this setting. In this section, we demonstrate with the following example that the algorithm in Table \ref{t:smcsamplers}B can expose deficiencies of partially observed, nonlinear dynamic models.

The disease dynamics of influenza infecting humans in temperate regions are characterized by explosive, seasonal epidemics during the winter months and marked, irregular variation across consecutive seasons. To understand the epidemiology of seasonal influenza, various complex non-linear dynamic models that track the number of susceptible (S), infected (I) and recovered/ immune (R) individuals in a population have been considered \citep[]{Earn2002}. Much of this work is motivated by the fact that simple models that assume gradual loss of immunity to reinfection do not fully describe observed disease patterns. We consider a stochastic SIRS model defined by the transmission rates
\begin{equation}\label{e:SIRS}
	\begin{split}
			&\frac{dS}{dt}=\:\mu(N-S)-\beta_t\frac{S}{N}I+\gamma(N-S-I)\\
			&\frac{dI}{dt}=\:\beta_t\frac{S}{N}I-(\mu+\nu)I,
	\end{split}
\end{equation}
where $\mu$ is the birth/death rate, $N$ is the finite population size, $\nu$ is the recovery rate, $\gamma$ the rate by which immunity is lost and $\beta_t$ the transmission rate that is further assumed to vary seasonally as
\begin{equation*}
\beta_t= \beta\big(1+s \sin(2\pi t)\big).
\end{equation*}
In practice, we account for long-term demographic trends (thus, $N$ and $\mu$ are fixed), and reparameterize $\beta$, $\nu$, $\gamma$ in terms of the basic reproductive number $R_0= \beta_t / (\nu+\mu)$, the average duration of infection per day, $D= 1/\nu$, and the average duration of immunity per year, $\Gamma= [365\gamma]^{-1}$; see the Appendix for further details. We fit and contrast \eqref{e:SIRS} to weekly Dutch influenza-like illness data ($x_0$) of influenza A (H3N2) that was collected between $1968-99$ (http://www.nivel.nl/griep/). Influenza-like illness data is subject to fluctuating reporting biases. We use a Poisson observation model that accounts for unknown reporting biases $\rho$ and known seasonal fluctuations in reporting fidelity $f_t$ (see Appendix),
\begin{equation}\label{e:SIRSobs}
\frac{dx}{dt}=\:\rho f_t^{-1} I,
\end{equation}
yielding five unknown parameters $\theta= (R_0, D, \Gamma, s, \rho)$ in total. The prior densities are $R_0\sim \mathcal{U}(1,20)$, $\Gamma\sim\mathcal{U}(1,160)$, $s\sim\mathcal{U}(0.075, 0.6)$, $\rho\sim\mathcal{U}(0.04,0.4)$ (uninformative), and $D\sim\mathcal{U}(2.2, 2.8)$ (informative) \citep[]{Gupta1998}. We use the standard Euler multinomial scheme \citep[]{Tian2004} to simulate from (\ref{e:SIRS}-\ref{e:SIRSobs}).

To compare the observed influenza-like illness counts $x_0$ to simulated data $x$, we use five summary statistics that reflect the characteristic dynamic features of influenza A (H3N2). Interannual seasonal variation is primarily reflected by the cumulative distribution and autocorrelation in peak differences (\CFDPK, \ACFDPK). The explosiveness of seasonal epidemics is reflected by the average duration of an epidemic at or above half its peak size (\MWIDTH), and we query overall magnitude with the cumulative distribution of peak incidence (\CPK) and the average annual attack rate (\MATTR). Here, annual attack rates are computed directly as the ratio of cumulative influenza-like illness counts in a winter season against average population size in that season; see the Appendix for further details.

The sequential algorithm \smcABCmu\ (Table \ref{t:smcsamplers}B) readily detects several shortcomings of the nonlinear stochastic SIRS model, while fitting it simultaneously to the data. Here, we actually seeded \smcABCmu\ with a few samples from algorithm \mhABCmu\ taken after burn-in, rather than seeding from the prior density. As we discuss later, this hybrid approach (hybrid\ABCmu) often improves overall efficiency. Our sequential sampler was then run for two iterations at the final ABC thresholds with $1000$ particles. We find that model (\ref{e:SIRS}-\ref{e:SIRSobs}) produces disease dynamics with too regular and too strong interannual variation. In Figure \ref{f:SIRSdiscrepancies}, we display a representative particle sampled at the last ABC iteration and two-dimensional projections of the estimated, five-dimensional ABC error density \eqref{e:ABCmuerrordensity}. 

\begin{table*}[tbp]
\centering
{\footnotesize
\begin{tabular}{llll}
	\hline
 				& burn-in 	& ESS/1000 & \#sim/ESS  \\ 
	\hline
	\mhABCmu 	& 4639 [3963, 5178]		& 60 [49, 75] 		& 859 [705, 1186]\\ 
	\smcABCmu 	& 24286 [21117, 26481]	& 125 [34, 184]		& 557 [374, 1834]\\ 
	hybrid\ABCmu	& 19263 [18884, 19605]	& 117 [41, 200]		& 363 [194, 968]\\ 
	\hline
\end{tabular}
}
\caption{Case study to analyze the performance of algorithms \mhABCmu, \smcABCmu, and the sequential importance sampler that is seeded with \mhABCmu\ (hybrid\ABCmu) on the example in Section \ref{s:dynamic}. The first column gives the burn-in of the algorithms under investigation. The second column reports effective sample size per 1000 samples from \eqref{e:ABCmutargetdensity} after burn-in. The third column gives the number of total simulations, including burn-in, per effective sample.}\label{t:casestudy}
\end{table*}

We also compared the numerical efficiency of algorithms \mhABCmu, \smcABCmu, as well as the hybrid approach in
terms of simulation effort per effectively independent sample from the target density
\eqref{e:ABCmutargetdensity}. Here, we present results on the SIRS model (\ref{e:SIRS}-\ref{e:SIRSobs}).
Algorithm \mhABCmu\ was run with annealing on the ABC threshold and the variance of the proposal density. Four
chains were generated in parallel. For comparison, burn-in of \mhABCmu\ is here the number of iterations to 
anneal to the final ABC threshold, summed across the four chains. The effective sample size (ESS) was computed by the method of \cite{Sokal1989}.
Algorithm \smcABCmu\ was run with 1000 particles for five iterations with annealing on the ABC threshold and the variance of the proposal density. Algorithm hybrid\ABCmu\ used 100 thinned samples from
500 iterations of four Markov chains after burn-in to seed a sequential importance sampler with 1000 particles
for two more iterations at the final ABC threshold; these configurations gave best results in terms of the total number
of simulations per effective sample (\#sim/ESS). For both \smcABCmu\ and hybrid\ABCmu, burn-in is the number 
of simulations to reach the final iteration of the sequential importance sampler, which also 
counts simulations from \mhABCmu. Here, ESS was calculated from the particle weights of the final iteration.
The results in Table \ref{t:casestudy} are obtained from
100 replications of all algorithms, and may not be directly comparable because the methods to compute effective
samples sizes differ. Comparing overall simulation effort (third column) suggests to us that the hybrid sampler
performs best as it combines rapid convergence with an efficient method to generate effectively independent
samples from the target density (first and second column). 



\section{Application of \ABCmu\ to population genetics}\label{s:phylotrees}
The ABC errors make possible to contrast a model against the real data in absolute terms. While this enables to successively refine any given model, there is, currently, no general approach to compare two models based on these errors. We illustrate the interplay of model checking and model comparison on a simple population genetics example where the true Bayes factor can be numerically estimated.

\begin{figure}[h!]
  \centering
  \includegraphics[width=.5\textwidth]{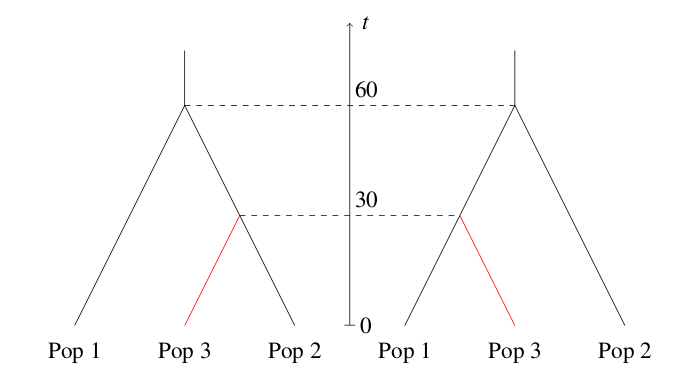}
  \caption{The two evolutionary scenarios considered in Section \ref{s:phylotrees}.
    A coalescent process evolves under time over each branch of the two scenarios. 
    \label{fig:scenarios} \textit{Left:} Model~Div23, where Population~3 diverged from Population~2 at time $t=30$ 
    \textit{Right:} Model~Div13, where Population~3 diverged from Population~1 at time $t=30$}
\end{figure}

Population genetics seek to infer aspects of the evolutionary history of a population of related individuals from genetic data. We assume that the
different populations evolve from a common ancestral population and that a tree topology encodes this history. The dynamics of changing genotype frequencies in populations that comply with the tree topology can often be represented by the Coalescent process, and methods have been developed to infer the associated process parameters from a small sample of genetic data of these populations \citep[]{crow, Kingman}.

\begin{figure*}[tbp]
	\includegraphics[type=pdf,ext=.pdf,read=.pdf,width=\textwidth]{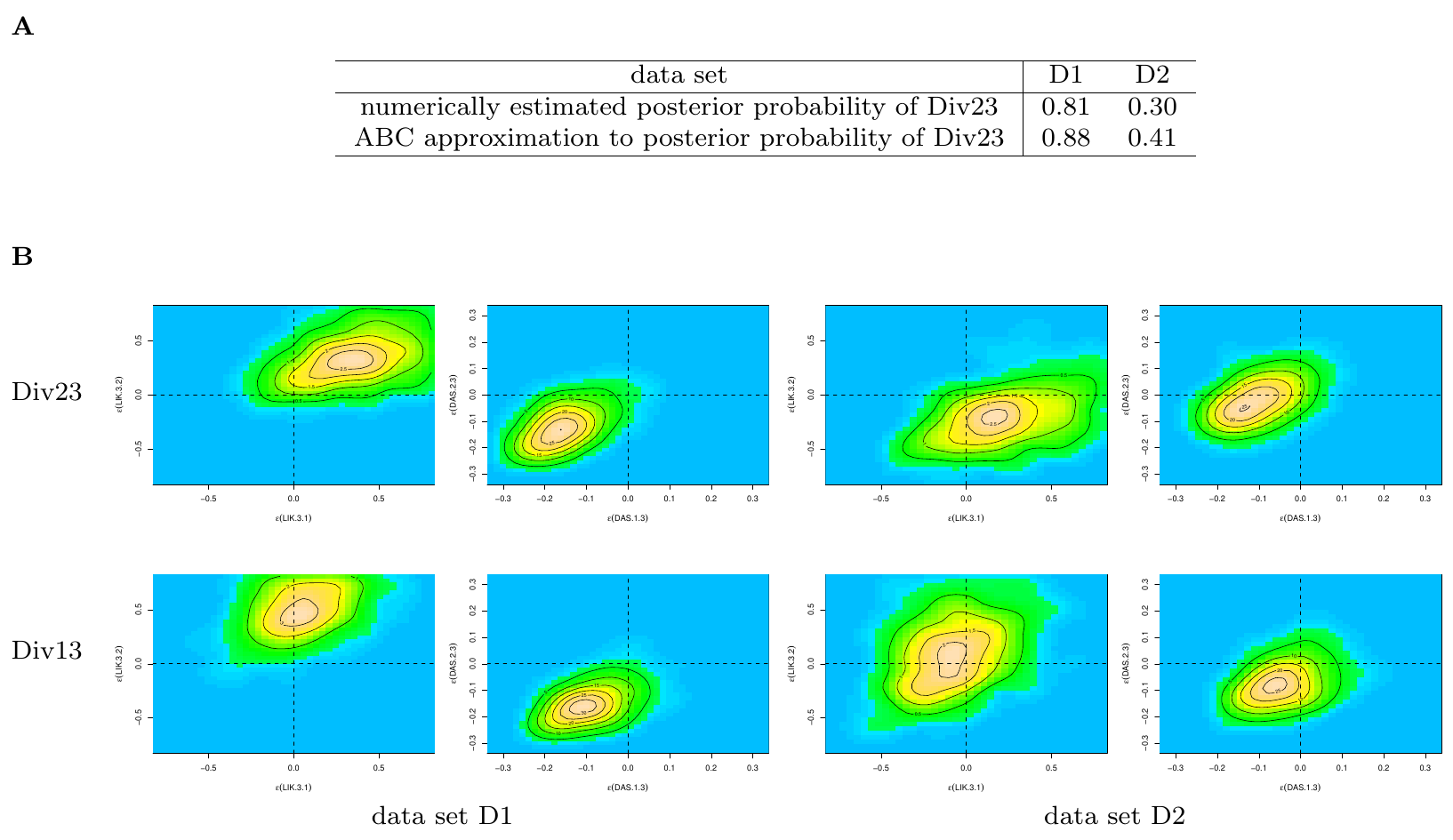}		
	\caption{%
The interplay of model assessment via ABC errors and model comparison via the Bayes factor on the phylogenetic simulation study in Section \ref{s:phylotrees}. (A) Estimates of the posterior probability of model Div23, computed with importance sampling and ABC, under equal prior probabilities for each model. Comparing both models relative to each other, Div23 is favored under data set D1 and Div13 under D2. (B) Estimates of some two-dimensional ABC error density for both models and both data sets. Comparing each models in absolute terms against the data, only model Div13 matches data set D2 reasonably well.
}\label{f:popgen1}
\end{figure*}

We consider here the case of three populations having evolved according to one of the two models described in
Figure~\ref{fig:scenarios}. Under both models, some individuals were genotyped from each population at time $t=0$.
Populations 1 and 2 diverged at a known date ($t=60$), and the deepest divergence event in the tree occurred at
$t=30$. Each sample from one of the three populations is composed of $15$ diploid individuals, which have been
genotyped at 5 independant microsatellite loci (different genotypes differ in the size of the repeated region).
Evolution of these loci over time follows the above model, i.e., when a mutation occurs, the length of the
sequence is increased or decreased by the length of the short repeated motif \citep{Ohta1973}.
The per capita mutation rate was set to $0.005$ for all five loci. We only infer the effective population size
$N_e$, assuming homogeneity across branches of the different scenarios. The prior density of $N_e$ was set to
$\mathcal{U}(2,150)$. This setup is quite simple in order to allow computation of the true posterior density
with importance sampling techniques \citep[combining techniques based on][]{stephens:donnelly:2000}.

To perform ABC analysis, the data are summarized through $24$ statistics. Some of these numerical summaries are
sensitive to the distance on the tree topology of two out of the three population samples taken at $t=0$.  We consider
Wright's measure of population heterogeneity between two populations (FST) \citep[]{weir:cockerham:1994}, the negative
log-likelihood (LIK) that samples from one population actually come from another population \citep[]{rannala:mountain:1997},
as well as the shared allele distance between two populations (DAS) \citep[]{chakraborty:jin:1993}. We write FST.$a$.$b$ when
FST is computed between samples corresponding to populations $a$ and $b$, and likewise for DAS and LIK. This gives all
in all six summary statistics. While FST and LIK are positively correlated with genetic divergence, DAS is negatively
correlated.

To compare different methods for ABC model choice in a controlled setting, we simulated two data sets under
model Div23 with $N_e=75$ and applied algorithm rej\ABCmu; see the Appendix.  Both models are similar enough so
that the evidence for model Div13 can be larger than the evidence for Div23.  Figure \ref{f:popgen1}A reports
importance sampling estimates of the true Bayes factor as well as ABC approximations thereof. The former are
computed by importance sampling approximations to both marginal likelihoods, while the later are
computed as ratios of frequencies of acceptances of simulations from both models, see \cite{Robert2011}   for details.
\cite{Robert2011} demonstrated that the ABC approximation of the Bayes factor does not converge to the
(numerically estimated) Bayes factor with increasing sample size and/or increasing computation runtimes.
Considering the first data set (D1), the evidence for model Div23 is larger when compared to the evidence for
Div13. For the second data set (D2), the situation is reversed, reflecting chance events with which the
simulated data sets were generated. The ABC errors in Figure \ref{f:popgen1}B reveal that, only, Div13
matches data set D2 reasonably well. Model Div13 shows larger discrepancies on this data set, which is in
agreement with the Bayes factor. Turning to data set D1, the ABC errors show that none of the fitted models
are consistent with the data, although the errors are slightly smaller under model Div23, again in agreement
with the Bayes factor.


\section{Discussion}\label{s:discuss}

We showed how existing ABC algorithms can be modified to aid targeted model refinement on the fly, and
illustrated these algorithms on examples from evolutionary biology and dynamic systems. Underlying this
workhorse machinery is a simple re-interpretation of ABC as a particular data augmentation
method on error space, and the presented algorithms are an immediate consequence of this reinterpretation.
Previously, we recognized the utility of the ABC errors as (unknown but estimable) compound random variables
that may reflect both stochastic fluctuations and systematic discrepancies between a model and the data
\citep[]{Ratmann2009}. There are two elementary points to make. The joint distribution of ABC errors
\eqref{e:ABCmuerrordensity} faithfully reflects dependencies among different aspects of the full,
intractable data. Indeed, the product ABC kernel that is used here does not confound the relation 
between multiple, lower-dimensional projections \eqref{e:Kdimerror} on error space. Thus, each dimension of the
ABC error density retains an intrinsic meaning and can be used to diagnose specific model deficiencies. Second,
the ABC errors are already computed by existing ABC algorithms (see Tables
\ref{t:rejectionsamplers}-\ref{t:smcsamplers}) and there is no further computational cost associated for the
purpose of model assessment.

The ABC error density \eqref{e:ABCmuerrordensity} provides an exploratory rather decisional tool for targeted,
iterative model refinement. One obstacle in using \eqref{e:ABCmuerrordensity} more formally for the purpose of
model comparison is that the ABC errors have an intrinsic scale that is model-dependent, and are often not
directly comparable. However, as there is no direct connection or convergence of approximations to the Bayes
factor to the true Bayes factor between two alternative models \citep[]{Robert2011}, there is renewed interest in
harnessing the information provided by \eqref{e:ABCmuerrordensity} for ABC model choice. Approximations to the
deviance information criterion \citep[]{Francois2011} provide an overall measure of goodness of fit
that complements the approach taken here. Clearly, both methods are often sensitive to the choice of the ABC
thresholds $\tau$ in the same way as ABC parameter inference is sensitive to $\tau$, and caution is warranted.
Further work on the selection of those thresholds is clearly needed, along the lines drafted by
\cite{Blum2009b} and \cite{Pudlo2011}.

In general, the ABC error density \eqref{e:ABCmuerrordensity} does not necessarily uncover existing model
deficiencies. 
It may be that the level of information contained in the data is too low to eliminate an inadequate model. An ideal setting is when both
summaries and ABC thresholds can be chosen adequately in order to expose those deficiencies, but this often is
an unachievable goal, if only for computational reasons. 
At this stage, we can only make the following simple
recommendations. For complex models in many real-world applications, the errors $\beps$ are typically dependent
and possibly antagonistic in that they change in different directions as $\theta$ is changed. In case of model
mismatch, we thus expect an irreconcilable conflict between components of $\varepsilon_{1:K}$
\citep[]{Robert2009PNAS,Ratmann2010}. This multi-dimensional perspective leverages the power of the ABC error
density in uncovering existing discrepancies considerably. More precisely, it follows from the results in
Section \ref{s:section1} that the expected ABC error vector
\begin{equation*}
\E_{\pi_\tau}(\beps|x_0)=\Big(\int \varepsilon_k \abcmumargtarget d\beps\Big)_{1:K}
\end{equation*}
is in each dimension weighted according to the ``mutual constraints" $\prod_k \abcK\big(\varepsilon_k;\tau_k\big)$:
\begin{equation*}
\begin{split}
&\int \varepsilon_k \abcmumargtarget d\beps \\
&=\negthinspace\int\negthinspace\negthinspace\rho_k\big(S_k(x),S_k(x_0)\big)\negthinspace\prod_j \negthinspace\abcK\Big(\negthinspace\rho_j \big(S_j (x),S_j (x_0)\big);\tau_j\negthinspace\Big)\pi(dx).
\end{split}
\end{equation*}
From a theoretical point, there is no possibility of conflict between summary errors whenever 
the respective summaries are independent of each other,
\begin{equation*}
\begin{split}
&\int \varepsilon_k \abcmumargtarget d\beps\\
&=\negthinspace\int\negthinspace\rho_k\big(S_k(x),S_k(x_0)\big)\abcK\Big(\rho_k\big(S_k(x),S_k(x_0)\big);\tau_k\Big)\pi(dx);
\end{split}
\end{equation*}
see the Appendix. As before, conflict is sensitive to the choice of the tolerance $\tau$ and vanishes as
$\tau\rightarrow\infty$. Crucially, conflict may emerge between as few as two co-dependent summary errors. We
thus often do not require a set of sufficient summaries to be able to detect existing model discrepancies.
Note, however, that efforts to orthogonalize a set of summaries \citep[]{Nunes2010} may diminish chances to
uncover model deficiency.

Based on our experience with the algorithms presented in Tables \ref{t:mhsamplers}-\ref{t:smcsamplers}, we
further recommend to combine \mhABCmu\ and \smcABCmu\ to sample from \eqref{e:ABCmutargetdensity}. Intuitively,
\mhABCmu\ updates only a single particle in relation to its previous value, and may lead to rapid convergence
to the target density. Techniques aiding rapid convergence are discussed in \cite{Ratmann2007}. As the
Metropolis-Hastings sampler might become stuck \cite{Sisson2007}, we run multiple chains in parallel until
shortly after burn-in. Our acceptance probabilities of \mhABCmu\ are typically below 5\%, hence the generated
Markov chains are highly autocorrelated. In our case study presented in Table \ref{t:casestudy}, we find that
\smcABCmu, seeded with samples from \mhABCmu, may quickly replenish effective sample sizes because it also
proposes ancestor indices. We present a small case study in Table \ref{t:casestudy}. There are perhaps two main
caveats with this hybrid approach. First, all Markov chains generated by \mhABCmu\ might get stuck. However,
this is a generic feature with all MCMC implementations that can be generically attenuated by the annealing
nature of \ABCmu\ (via the choice of the $\tau_{nk}$'s). Second, the acceptance probabilities in the rejection
control step of \smcABCmu\ can be significantly lower than those of \mhABCmu\ because the latter are not
computed in relation to previous error magnitudes. 

\section{Conclusion}
We presented methods and algorithms for exploratory model assessment when the likelihood is computationally intractable. To us, the advantages of this approach towards ABC model choice 
are that sufficient summaries are often not required to detect existing discrepancies between a model and the data, and that the algorithms incur no or little extra computational cost as compared to 
standard ABC methods. Perhaps, the main shortcomings of this approach may be the scale dependency of the ABC errors and their sensitivity to the ABC threshold. 
We believe that, in addition to investigating the effects of the ABC approximation to standard Bayesian machinery such as the Bayes factor or the deviance information criterion, exploiting the particular properties of that approximation may provide complementary, useful tools for ABC model choice.


%
%

\appendix
\section{Proofs of sections \ref{s:section1} and \ref{s:discuss}}
We first derive the marginal density of \eqref{e:ABCmutargetdensity} in $\theta$. By construction, it is
\begin{equation*}
\begin{split}
\pi^\text{ABC}(\theta|x_0) &\propto \pi(\theta) \int\abcK(\beps;\tau_k)\:\xi_{x_0,\theta}(d\beps)\\
&=\pi(\theta)\int\abcK\Big(\Big(\rho_k\big(S_k(x),S_k(x_0)\big);\tau_k\Big)_{1:K}\Big)\:f(dx|\theta).
\end{split}
\end{equation*}
Consequently, the marginal target densities $\abcmargtarget$ of algorithms rejABC and rej\ABCmu\ coincide if
\begin{equation*}
\abcK\big(\rho(\Sset(x),\Sset(x_0);\tau\big)\propto\prod_k \abcK\big(\rho_k(S_k(x),S_k(x_0);\tau_k\big)
\end{equation*} 
(We conjecture this property only happens for the normal kernel $\abcK$ and the Euclidean distance $\rho$, as
well as for the indicator kernel and the Manhattan distance.)
To establish \eqref{e:ABCmuerrordensity}, consider the following error measure on the (same) associated Borel
image $\sigma$-algebra under the ABC projection \eqref{e:Kdimerror},
\begin{equation*}
\begin{split}
&\pi_{x_0}(E_1\times\dotsc\times E_K)=\quad\int\pi(\theta)\xi_{x_0,\theta}(E_1\times\dotsc\times E_K)\:d\theta\\
&\quad=\iint_{\mathcal{X}}\pi(\theta)\:\Ind\Big\{x\in\xi_{x_0}^{-1}(E_1\times\dotsc\times E_K)\Big\}\:f(dx|\theta)\:d\theta\\
&\quad=\int_{\mathcal{X}}\Ind\Big\{x\in\xi_{x_0}^{-1}(E_1\times\dotsc\times E_K)\Big\}\bigg[\int\pi(\theta)\:f(dx|\theta)\:d\theta\bigg]\\
&\quad=\int_{\mathcal{X}}\Ind\Big\{x\in\xi_{x_0}^{-1}(E_1\times\dotsc\times E_K)\Big\}\:\pi(dx).
\end{split}
\end{equation*} 
Here, $\pi(dx)$ is the prior predictive distribution of the data \citep[]{Box1980}, and $\pi_{x_0}(d\beps)$ can
thus be interpreted as the prior predictive $K$-dimensional error measure under the ABC projection.  We assume
that $\pi_{x_0}$ admits a density (also denoted by) $\pi_{x_0}\colon\R^K\to\R^+_0,\:\beps\to\pi_{x_0}(\beps)$
with respect to the same measure as \eqref{e:xiK}, and define the algorithm outcome as
\begin{equation*}
\abcmumargtarget\propto \prod_k \abcK\big(\varepsilon_k;\tau_k\big)\:\pi_{x_0}(\beps).
\end{equation*}
We previously derived the same relations under a Riemann interpretation of the integrals involved
\citep[]{Ratmann2009}. The Lebesgue approach presented here is much less convoluted.

Finally, we establish the intuitive result that conflict cannot emerge between independent summary errors. If the summary errors in ABC are independent of each other,
\begin{equation*}
\mathbb{P}_{\theta,x_0}\Big(E_1,\dotsc,E_K\Big)=\int\Ind\Big\{x\in \xi_{x_0}^{-1}(E_1\times\dotsc\times E_K)\Big\}\:f(dx | \theta)
\end{equation*}
factorizes componentwise.  Likewise, the prior predictive $K$-dimensional error density $\pi_{x_0}(\beps)$ admits the factorization $\pi_{x_0}(\beps)=\prod_{k=1}^K\pi_{k,x_0}(\varepsilon_k)$.
Since the ABC kernel is here assumed to factorize as well, we obtain 
\begin{align*}
&\abcmumargtarget\\
&\quad=\prod_{k=1}^K\pi_{k,x_0}(\varepsilon_k)\abcK(\varepsilon_k;\tau_k)\:\:\bigg/\\
&\hspace{1cm}\idotsint \prod_{k=1}^K\pi_{k,x_0}(\beps)\abcK(\varepsilon_k;\tau_k)\:d\varepsilon_1\dotsc d\varepsilon_K\\
&\quad=\prod_{k=1}^K\pi_{k,x_0}(\varepsilon_k)\abcK(\varepsilon_k;\tau_k)\:\:\bigg/\:\prod_{k=1}^K\int \pi_{k,x_0}(\varepsilon_k)\abcK(\varepsilon_k;\tau_k)\:d\varepsilon_k\\
&\quad= \prod_{k=1}^K \pi_{k,\tau_k}(\varepsilon_k|x_0).
\end{align*}
Therefore, the $k$th component of the vector-valued mean ABC error $\E_{\pi_\tau}(\beps|x_0)$ collapses to
\begin{equation*}
\begin{split}
&\int \varepsilon_k\prod_{k=1}^K\pi_{k,\tau_k}(\varepsilon_k|x_0)\:d\beps\\
&\quad=\int\varepsilon_k\:\pi_{k,\tau_k}(\varepsilon_k|x_0) \bigg[\idotsint\prod_{j\in -k}\pi_{j,\tau_j}(\varepsilon_j|x_0)\:d\varepsilon_{-k}\bigg]d\varepsilon_k\\
&\quad=\int\varepsilon_k\:\pi_{k,\tau_k}(\varepsilon_k|x_0)d\varepsilon_k\\
&\quad=\int\rho_k\big(S_k(x),S_k(x_0)\big)\:\pi_{\varepsilon_k}\Big(\rho_k\big(S_k(x),S_k(x_0)\big);\tau_k\Big)\:\pi(dx).
\end{split}
\end{equation*}

\section{Simulation from the network models}
Each of the considered network evolution models defines a discrete-state discrete time Markov chain of growing networks. For network data of some organism, we grow a network according to model-specific transition probabilities until the number of nodes in the network equals the number of genes in the genome of that organism. Transition probabilites are implicitly defined by the following probabilistic rules. The preferential attachment (PA) model adds a new node to an existing node with probability that is proportional to the node degree of the existing node. In the duplication-divergence model (DD), a parent node is randomly chosen and its edges are duplicated. For each parental edge, the parental and duplicated one are then lost with probability $\pDiv$ each, but not both; moreover, at least one link is retained to any node. The parent node may be attached to its child with probability $\pAttach$. The mixture model DD+PA either performs PA with probability $\alpha$, or DD with probability $1-\alpha$. Model DD+LNK+PA is a mixture of PA, DD with $\pAttach=0$, link addition and deletion. Link addition (deletion) proceeds by choosing a node randomly, and attaching it preferentially to another node (deleting it preferentially from its interaction partners). Unnormalized mixture weights are calculated as follows. For duplication-divergence, the rate $\rDup$ is multiplied by the order of the current network; for link addition, the rate $\rLnkAdd$ is multiplied by $\binom{\text{Order}}{2}-\text{Size}$; for link deletion, the unnormalized weight of link addition is multiplied by $\rLnkDel$. Preferential attachment occurs at a constant frequency $\alpha$, and the weights of duplication, link addition and link deletion are normalized so that their sum equals $1-\alpha$. Each of the components is chosen according to the normalized weights.

We consider here two alternative observation models to account for missing data. Denote the number of proteins in the observed data set $x_0$ with $n(x_0)$, the number of bait proteins with $n_{\rm bait}(x_0)\leq n(x_0)$, the number of prey proteins with $n_{\rm prey}(x_0)\leq n(x_0)$, the number of links with $m(x_0)$ and the fully simulated network with $\tilde{x}$. Link subsampling (L): Set $x$ to be empty and repeat until the number of links in $x$ is equal or larger than $m(x_0)$, or no more links can be added: pick a link $(u,v)$ at random and without replacement from $\tilde{x}$ and add the nodes $u$, $v$ and the link $(u,v)$ to $x$. Bait-prey subsampling (BP): Create two random lists of bait and prey proteins. Set $x$ to be empty and repeat until the number of baits $n_{\rm bait}(x)$ equals  $n_{\rm bait}(x_0)$ and the number of preys $n_{\rm prey}(x)$ is equal or larger than $n_{\rm prey}(x_0)$, or no more links can be added: pick a link $(u,v)$ at random and without replacement from $\tilde{x}$ such that $u$ is in the bait list and $v$ is in the prey list, mark $u$ as a bait and $v$ as a prey protein and add the nodes $u$, $v$ and the link $(u,v)$ to $x$. 

\section{Simulation from the stochastic SIRS model}
Assuming that the infinitesimal probability of either simultaneous or multiple transitions between compartments is negligible, the transmission rates (\ref{e:SIRS}-\ref{e:SIRSobs}) define a finite-state continuous time Markov chain that accounts for demographic stochasticity. To adjust for long-term demographic trends in the Netherlands, we fix $N$ to historical population data obtained from (http://statline.cbs.nl/statweb/), and set $1/\mu= 80$. To avoid stochastic extinction, we add a small, constant number of infected visitors $I_v$ which can be interpreted as the average number of infected travelers that are visiting the Netherlands at any day. The first equation in \eqref{e:SIRS} is thus
\begin{equation*}
\frac{dS}{dt}=\:\mu(N-S)-\beta_t\frac{S}{N}(I+I_v)+\gamma(N-S-I).
\end{equation*}
The value $I_v$ is set to the expected number of infected travelers for a given model parameterization. More precisely, we estimate the average total number of international travelers in the Netherlands at $5$m/yr from available tourist information (http://statline.cbs.nl/statweb/) and set $I_v$ the proportion of international travelers that would be infected at any day at endemic equilibrium,
\begin{equation*}
I_v= \frac{\mu+\gamma}{\mu+\gamma+\nu}(1-1/R_0)\times 5\times 10^6/365.
\end{equation*}
The Euler-Maryuama algorithm is incremented by $0.5$ days. 

The observation model used here accounts for seasonal differences in the true positive rate $f_t$ of reported influenza-like illness cases that are subsequently confirmed as true influenza cases with virological analyses.  Effectively, our model inflates simulated summer incidence to larger values, because typically less than 5\% of reported influenza-like illness cases are confirmed during the summer period. We do not have access to the known true positive rate in the Netherlands. Inspecting available incidence data and true positive rates from the U.S. between 1997-2008 (http://www.cdc.gov/flu/weekly /fluactivitysurv.htm), we find it is possible to predict $f_t$ in one season reasonably accurately from the timing of peak incidence in the same season.  

\section{Summaries for the stochastic SIRS model}
We investigated a much larger number of candidate summaries, and found that the set (\CFDPK, \ACFDPK,\MWIDTH,\CPK,\MATTR) is sufficient to expose model deficiencies against the observed data and to estimate model parameters accurately in simulation studies. We used the following distance functions for each of the summaries. For \CFDPK, we compute the Cramer-von-Mises test statistic. For \ACFDPK, we compute the log ratio of the observed and simulated autocorrelation at lag 2 to reproduce influenza A (H3N2)'s weak biennial oscillation. For \MWIDTH\ and \MATTR, we compute the log ratio of the observed and simulated means. The summary \CPK\ is subject to considerable volatility upon re-simulation, which precluded the use of the the Cramer-von-Mises test statistic. Most of the parameter space results in strongly bimodal peak distributions. To identify the small parameter space for which more gradual peak distributions are obtained, we found it most efficient to query the discrepancies between the two cumulative distribution functions at peak size 200 and 400 in terms of the log ratio. Here, we use log ratios and tail area probabilites rather than difference functions so that the resulting ABC errors have some intrinsic interpretability.

\section{Simulation from the population genetic models}
We used the DIYABC software \citep[see][]{cornuet:santos:beaumont:etal:2008} to produce simulations from model Div13 and
Div23. The genotypes of the simulated samples are drawn independently for each locus. Following the population history of
the model and assuming no natural selection, the gene genealogy (which can be displayed as a dendrogram rooted at the most
recent common ancestor) is given by time-continuous Coalescent process. Coalescence rate in the genealogy is governed by
the effective population size $N_{e}$: in a population, if $k$ ancestors of the sample remain at a given time, a
coalescent event (joining two branches of the dendrogram chosen at random) occurs after an exponential time with rate
$k(k-1)\big{/}(2N_{e})$.  Hence the number of branches over (backward) time in a population evolves according to a pure death
Markov process, jumping from state $k$ to state $k-1$ with rate $k(k-1)\big{/}(2N_{e})$. Conditionally on the genealogy
of this locus, mutation rate ($\mu$) over each branch of the dendogram is assumed to be $0.005$. We can then genotype
the simulated sample, starting from the genotype of the most recent common ancestor and respecting the mutation model
described in Section~\ref{s:phylotrees}.

We sum up the whole simulated data set with the $24$ usual summary statistics given in \citet{Robert2011}.  In the
simple situation considered here, it is possible to apply the rejection algorithm, and in fact, we here re-evaluated the
same pseudo data sets that were also used in \citep[]{Robert2011}. The acceptance function $\kappa$, see
Table~\ref{t:rejectionsamplers}, is a product of $24$ one-dimensional indicator functions centered around $0$. And the
tolerances $\tau_k$ are tuned so that we keep in Model Div23 about $1000$ simulated data sets from the $10^6$ simulated
data sets of the reference table. Our conclusions on model assessment rely the error distribution of six summary
statistics, which are correlated with the distances in genetic variation between populations.


\begin{acknowledgements}
We would like to thank Jean-Marie Cornuet for the simulations with the DIYABC software that were used in
Section~\ref{s:phylotrees}, and Christophe Andrieu for stimulating discussions.
\end{acknowledgements}

\bibliographystyle{spbasic}      

\begin{thebibliography}{50}
\providecommand{\natexlab}[1]{#1}
\providecommand{\url}[1]{{#1}}
\providecommand{\urlprefix}{URL }
\expandafter\ifx\csname urlstyle\endcsname\relax
  \providecommand{\doi}[1]{DOI~\discretionary{}{}{}#1}\else
  \providecommand{\doi}{DOI~\discretionary{}{}{}\begingroup
  \urlstyle{rm}\Url}\fi
\providecommand{\eprint}[2][]{\url{#2}}

\bibitem[{Beaumont(2010)}]{Beaumont2010}
Beaumont MA (2010) Approximate Bayesian computation in evolution and ecology.
  Ann Rev Ecol Ev and Syst 41(1):379--406

\bibitem[{Beaumont et~al(2010)Beaumont, Cornuet, Marin, and
  Robert}]{Beaumont2009}
Beaumont MA, Cornuet JM, Marin JM, Robert CP (2010) {Adaptivity for ABC
  algorithms: the ABC-PMC scheme}. Biometrika 96(4):983--990

\bibitem[{Blum(2009)}]{Blum2009b}
{Blum M (2009) Approximate Bayesian Computation: a non-parametric perspective.
  Tech. Rep. 0904.0635v4, arXiv
}

\bibitem[{Bortot et~al(2007)Bortot, Coles, and Sisson}]{Bortot2007}
Bortot P, Coles S, Sisson S (2007) Inference for stereological extremes. J Am
  Stat Ass 102(9):84--92

\bibitem[{Box(1980)}]{Box1980}
Box GEP (1980) Sampling and {B}ayes' inference in scientific modelling and
  robustness. J Roy Stat Soc A (General) 143(4):383--430

\bibitem[{Chakraborty and Jin(1993)}]{chakraborty:jin:1993}
Chakraborty R, Jin L (1993) A unified approach to study hypervariable
  polymorphisms: statistical considerations of determining relatedness and
  population distances. EXS 67:153--175

\bibitem[{Cornuet et~al(2008)Cornuet, Santos, Beaumont, Robert, Marin, Balding,
  Guillemaud, and Estoup}]{cornuet:santos:beaumont:etal:2008}
Cornuet JM, Santos F, Beaumont MA, Robert CP, Marin JM, Balding DJ, Guillemaud
  T, Estoup A (2008) Inferring population history with {DIYABC}: a
  user-friendly approach to {A}pproximate {B}ayesian {C}omputation.
  Bioinformatics 24(23):2713--2719

\bibitem[{Crow and Kimura(1970)}]{crow}
Crow JF and Kimura M (1970) An introduction to population genetics theory. Harper \& Row, Publishers

\bibitem[{Dean et~al(2011)Dean, Singh, and Peters}]{Dean2011}
Dean TA, Singh A S~Sand~Jasra, Peters G (2011) Parameter estimation for hidden
  markov models with intractable likelihoods. arXiv:11035399

\bibitem[{{Del Moral} et~al(2008){Del Moral}, Doucet, and Jasra}]{Moral2008}
{Del Moral} P, Doucet A, Jasra A (2008) An adaptive {Sequential Monte Carlo}
  method for {Approximate Bayesian Computation}. Tech. rep., University of
  Bordeaux, France,
  \urlprefix\url{http://stats.ma.ic.ac.uk/a/aj2/public\_html/papers/delmo ral\_doucet\_jasra\_smcabc.pdf}

\bibitem[{Drovandi et~al(2011)Drovandi, Pettitt, and Faddy}]{Drovandi2011}
Drovandi CC, Pettitt AN, Faddy MJ (2011) {Approximate Bayesian computation
  using indirect inference}. J Roy Stat Soc (C) 60(3):317--337

\bibitem[{Earn et~al(2002)Earn, Dushoff, and Levin}]{Earn2002}
Earn DJD, Dushoff J, Levin SA (2002) Ecology and evolution of the flu. Trends Ecol Evol 17(7):334--340

\bibitem[{{Fearnhead} and {Prangle}(2010)}]{Fernhead2011}
{Fearnhead} P, {Prangle} D (2010) {Constructing Summary Statistics for
  Approximate Bayesian Computation: Semi-automatic ABC}. ArXiv e-prints
  \eprint{1004.1112}

\bibitem[{{Francois} and {Laval}(2011)}]{Francois2011}
{Francois} O, {Laval} G (2011) {Deviance Information Criteria for Model
  Selection in Approximate Bayesian Computation}. ArXiv e-prints
  \eprint{1105.0269}

\bibitem[{Grelaud et~al(2009)Grelaud, Robert, Marin, Rodolphe, and
  Taly}]{grelaud2009}
Grelaud A, Robert CP, Marin JM, Rodolphe F, Taly JF (2009) 
  {Likelihood-free methods for model choice in Gibbs random fields}. Bayesian
  Analysis 4(2):317--336

\bibitem[{Gupta et~al(1998)Gupta, Ferguson, and Anderson}]{Gupta1998}
Gupta S, Ferguson N, Anderson R (1998) Chaos, persistence, and evolution of
  strain structure in antigenically diverse infectious agents. Science
  280(5365):912--915

\bibitem[{{Heggland} and {Frigessi}(2004)}]{Heggland2004}
{Heggland} K, {Frigessi} A (2004) Estimating functions in indirect inference. J
  Roy Stat Soc B 66:447--462

\bibitem[{Jaakkola and Jordan(2000)}]{Jaakkola2000}
Jaakkola TS, Jordan MI (2000) Bayesian parameter estimation via variational
  methods. Statistics and Computing 10:25--37

\bibitem[{Kingman(1982)}]{Kingman}
Kingman JFC (1982) The Coalescent. Stoch. Proc. and Their Applications 13: 235--248

\bibitem[{Liu and Chen(1998)}]{Liu1998}
Liu JS, Chen R (1998) {Sequential Monte Carlo Methods for Dynamic Systems}.
  J Am Stat 93:1032--1044

\bibitem[{MacKay(2002)}]{MacKay2002}
MacKay DJC (2002) Information Theory, Inference \& Learning Algorithms.
  Cambridge University Press

\bibitem[{Marin et~al(2011)Marin, Pudlo, Robert, and Ryder}]{Pudlo2011}
Marin JM, Pudlo P, Robert CP, Ryder R (2011) Approximate Bayesian Computational
  methods. ArXiv e-prints \eprint{11010955}

\bibitem[{{Marjoram} et~al(2003){Marjoram}, {Molitor}, {Plagnol}, and
  {Tavar\'e}}]{Marjoram2003}
{Marjoram} P, {Molitor} J, {Plagnol} V, {Tavar\'e} S (2003) {M}arkov {C}hain
  {M}onte {C}arlo without likelihoods. Proc Natl Acad Sci USA
  100(26):15,324--15,328

\bibitem[{Nunes and Balding(2010)}]{Nunes2010}
Nunes MA, Balding DJ (2010) {On optimal selection of summary statistics for
  approximate Bayesian computation.} Stat App Gen Mol Biol 9(1)

\bibitem[{Ohta and Kimura(1973)}]{Ohta1973} 
  Ohta T, Kimura M (1973) A model of mutation appropriate to estimate the number
  of electrophoretically detectable alleles in a finite population. Genetics
  Research 22:201--204

\bibitem[{{Pritchard} et~al(1999){Pritchard}, {Seielstad}, {Perez-Lezaun}, and
  {Feldman}}]{Pritchard1999}
{Pritchard} J, {Seielstad} M, {Perez-Lezaun} A, {Feldman} M (1999) {Population
  growth of human Y chromosomes: a study of Y chromosome microsatellites}. Mol
  Biol Evol 16:1791--1798

\bibitem[{Rannala and Mountain(1997)}]{rannala:mountain:1997}
Rannala B, Mountain J (1997) Detecting immigration by using multilocus
  genotypes. Proc Natl Acad Sci USA 94:9197--9201

\bibitem[{Ratmann et~al(2007)Ratmann, J\o\!\;rgensen, Hinkley, Stumpf,
  Richardson, and Wiuf}]{Ratmann2007}
Ratmann O, J\o\!\;rgensen O, Hinkley T, Stumpf MP, Richardson S, Wiuf C (2007)
  Using likelihood-free inference to compare evolutionary dynamics of the
  protein networks of {H.pylori} and {P.falciparum}. PLoS Comp Biol
  3(2007):e230

\bibitem[{Ratmann et~al(2009)Ratmann, Andrieu, Wiuf, and
  Richardson}]{Ratmann2009}
Ratmann O, Andrieu C, Wiuf C, Richardson S (2009) Model criticism based on
  likelihood-free inference, with an application to protein network evolution.
  Proc Natl Acad Sci USA 106(26):10,576--10,581

\bibitem[{Ratmann et~al(2010)Ratmann, Andrieu, Wiuf, and
  Richardson}]{Ratmann2010}
Ratmann O, Andrieu C, Wiuf C, Richardson S (2010) Reply to {R}obert et al.:
  {M}odel criticism informs model choice and model comparison. Proc Natl Acad
  Sci USA 107(3):E6--E7

\bibitem[{Robert et~al(2009)Robert, Mengersen, and Chen}]{Robert2009PNAS}
Robert CP, Mengersen KL, Chen C (2009) Letter: Model choice versus model
  criticism. Proc Natl Acad Sci USA

\bibitem[{{Robert} et~al(2011){Robert}, {Cornuet}, {Marin}, and
  {Pillai}}]{Robert2011}
{Robert} CP, {Cornuet} JM, {Marin} JM, {Pillai} N (2011) {Lack of confidence in
  ABC model choice}. ArXiv e-prints \eprint{1102.4432}

\bibitem[{{Sisson} et~al(2007){Sisson}, {Fan}, and {Tanaka}}]{Sisson2007}
{Sisson} SA, {Fan} Y, {Tanaka} MM (2007) Sequential {M}onte {C}arlo without
  likelihoods. Proc Natl Acad Sci USA 104:1760--1765

\bibitem[{Sokal(1989)}]{Sokal1989}
Sokal A (1989) {Monte Carlo} methods in statistical mechanics: foundations and
  new algorithms. Tech. rep., Department of Physics, New York University

\bibitem[{Stephens and Donnelly(2000)Stephens and Donnelly}]{stephens:donnelly:2000}
 Stephens D and Donnelly P (2000) Inference in population genetics (with discussion).  J. {R}oyal {S}tatist.
 {S}oc.  62:602--655.

\bibitem[{{Stumpf} et~al(2007){Stumpf}, {Kelly}, {Thorne}, and
  {Wiuf}}]{Stumpf2006a}
{Stumpf} MPH, {Kelly} W, {Thorne} T, {Wiuf} C (2007) Evolution at the system
  level: the natural history of protein interaction networks. Trends Ecol Evol
  22:366--373

\bibitem[{Tian and Burrage(2004)}]{Tian2004}
Tian T, Burrage K (2004) {Binomial leap methods for simulating stochastic
  chemical kinetics.} J {C}hem {P}hys 121(21):10,356--10,364,
  \doi{10.1063/1.1810475}

\bibitem[{Titz et~al(2008)Titz, Rajagopala, Goll, H\"auser, McKevitt, Palzkill,
  and Uetz}]{Titz2008}
Titz B, Rajagopala SV, Goll J, H\"auser R, McKevitt MT, Palzkill T, Uetz P
  (2008) The binary protein interactome of {Treponema pallidum} -- the
  {Syphilis} spirochete. PLoS One 3(5):e2292

\bibitem[{Toni et~al(2008)Toni, Welch, Strelkowa, Ipsen, and Stumpf}]{Toni2008}
Toni T, Welch D, Strelkowa N, Ipsen A, Stumpf MPH (2008) Approximate {Bayesian}
  computation scheme for parameter inference and model selection in dynamical
  systems. J Roy Soc Interface

\bibitem[{Weir and Cockerham(1984)}]{weir:cockerham:1994}
Weir BS, Cockerham CC (1984) Estimating f-statistics for the analysis of
  population structure. Evolution 38:1358--1370

\end{thebibliography}

\end{document}